\documentclass[aps,prd,twocolumn]{revtex4}
\usepackage{graphicx}
\begin{document}

\title{\bf Deconfinement and color superconductivity in cold neutron stars}
\author{G. Lugones}\email{lugones@df.unipi.it}
\affiliation{Dipartimento di Fisica ``Enrico Fermi'' Universit\`a
di Pisa and INFN Sezione di Pisa,\\ Largo Bruno Pontecorvo 3,
56127 Pisa, Italy}

\author{I. Bombaci}\email{bombaci@df.unipi.it}
\affiliation{Dipartimento di Fisica ``Enrico Fermi'' Universit\`a
di Pisa and INFN Sezione di Pisa,\\ Largo Bruno Pontecorvo 3,
56127 Pisa,  Italy}


\begin{abstract}
We study the deconfinement transition of hadronic matter into quark
matter in neutron star conditions in the light of color
superconductivity. Deconfinement is considered to be a first order
phase transition that conserves color and flavor. It gives a
short-lived {($\tau \sim \tau_{weak}$)}    transitory
colorless-quark-phase that is {\it not} in $\beta$-equilibrium. We
deduce the equations governing deconfinement when quark pairing is
allowed and find the regions of the parameter space (pairing gap
$\Delta$ versus bag constant $B$) where deconfinement is possible
inside cold neutron stars. We show that for a wide region of
($B,\Delta$) a pairing pattern is reachable within a strong
interaction timescale, and the resulting ``2SC-like'' phase is
preferred energetically to the unpaired phase. We also show that
although $\beta$-stable hybrid star configurations are known to be
possible for a wide region of the ($B,\Delta$)-space, many of these
configurations could not form in practice because deconfinement is
forbidden, i.e. the here studied non-$\beta$-stable
\emph{intermediate} state cannot be reached.
\end{abstract}


\maketitle

\section{Introduction}

A general feature of degenerate Fermi systems is that they become
unstable if there exist any attractive interaction at the Fermi
surface. As recognized by Bardeen, Cooper and Schrieffer (BCS)
\cite{Bardeen1957} this instability leads to the formation of a
condensate of Cooper pairs and the appearance of superconductivity.
In QCD any attractive quark-quark interaction will lead to pairing
and color superconductivity, a subject already addressed in the late
1970s and  early 1980s \cite{Barrois1977,BailinLove} which came back
a few years ago since the realization that the typical
superconducting gaps in quark matter may be larger than those
predicted in these early works ($\Delta$ as high as $\sim 100$ MeV)
\cite{newgap}. The phase diagram of QCD has been analyzed in the
light of color superconductivity and model calculations suggest that
the phase structure is very rich at high densities. Depending on the
number of flavors, the quark masses, the interaction channels, and
other variables  many possible $\beta$-stable color superconducting
phases of quark matter are possible
\cite{Alford2001,Huang0311155,Rajagopal2000,Ruester2004,Nardulli}.

There is at present some indication that the quark gluon plasma
might have been produced in laboratory \cite{Gyulassy2004}.
However, it is not yet established whether deconfinement happens
in nature in the high density, low temperature regime that is
relevant for neutron stars. Unfortunately, first principle
calculations are not available in this region of the QCD phase
diagram. In turn we shall base our analysis on phenomenological
considerations which could delineate at least a broad brush
picture of the physics involved. Matter in compact stars should be
electrically neutral and colorless in bulk.  Also, any equilibrium
configuration of such matter should remain in $\beta$-equilibrium.
Satisfying these requirements impose nontrivial relations between
the chemical potentials of different quarks. Moreover, such
relations substantially influence the pairing dynamics between
quarks, for instance, by suppressing some color superconducting
phases and by favoring others \cite{Shovkovy2004}.

At not very large densities, the appearance of strange quarks is
suppressed because of their finite mass and quark matter is
composed almost completely by quarks $u$ and $d$. Pairing between
them would be possible if their Fermi momenta are not very
different. The resulting pairing pattern is the so-called
two-flavor color superconductor (2SC) in which the up and down
quarks  form Cooper pairs in the color-antitriplet,
flavor-singlet, spin-zero channel. In the conventional picture of
the 2SC phase it is assumed that pairs  are formed by up-red
($u_r$) and down-green ($d_g$) quarks, as well as by up-green
($u_g$) and down-red ($d_r$) quarks. The other two quarks ($u_b$
and $d_b$) do not participate in pairing. A more recent analysis
shows that the ground state of dense up-and-down quark matter
under local and global charge neutrality conditions with
$\beta$-equilibrium has at least four possibilities: normal,
regular 2SC, gapless 2SC phases, and mixed phase composed of 2SC
phase and normal components \cite{Huang0311155}. A new interesting
feature of some of these phases is that pairing is allowed between
particles even in the case $\delta \mu > \Delta$
\cite{Huang0311155}.

At sufficiently large densities the value of the chemical
potential exceeds the mass of the strange quark $m_s$. Strange
quarks appear in the mixture, and Cooper pairing can happen
between up, down and strange quarks. Pairing involving strange
quarks is expected to exist if the resulting gaps ($\Delta_{us}$
and $\Delta_{ds}$) are larger than $\sim m_s^2 / (2 \mu)$, the
difference between the $u$ and $s$ Fermi momenta in the absence of
pairing. Depending on the value of the strange quark mass, as well
as other parameters in the theory, many different paired
configurations are possible. At very high densities it seems clear
that the color-flavor locked phase (CFL) is the ground state.
However, at not very large densities, it is possible that up and
down quarks form 2SC matter, while the strange quarks do not
participate in pairing, eventually forming a $<ss>$ condensate
with a much smaller gap.  Finally, even more exotic crystalline
phases  of 1SC quark matter have been analyzed.

To the best of our knowledge, all previous works about color
superconductivity in compact stars have dealt with matter in
$\beta$-equilibrium. This is the situation expected to appear in
strange stars or hybrid stars as soon as they settle in a stable
configuration. However, during the deconfinement transition in
neutron stars, matter is transitorily out of equilibrium with
respect to weak interactions. In fact, the transition from
$\beta$-stable hadron matter to quark matter in cold neutron stars
should occur trough a quantum nucleation process
\cite{Kagan,Alcock,Horvath1994,Grassi1997,IidaSato1998,Bombaci2004}.
Quantum fluctuations could form both virtual drops of unpaired quark
matter (hereafter the $Q^*_{unp}$ phase) or virtual drops of
color-superconducting quark matter ($Q^*_{\Delta}$ phase). In both
cases, the flavor content of the quark matter virtual drop must be
equal to that of the confined $\beta$-stable hadronic phase at the
same pressure (the central pressure of the hadronic star). In fact,
since quark deconfinement and quark-quark pairing are due to the
strong interaction, the oscillation time $\nu_0^{-1}$ of a virtual
quark droplet in the potential energy barrier separating the
hadronic from the quark phase, is of the same order of the strong
interaction characteristic time ($\tau_{strong} \sim  10^{-23}$ s).
The latter is many orders of magnitude smaller than the weak
interaction characteristic time ($\tau_{weak} \sim  10^{-8}$ s).
Thus, quark flavor must be conserved forming a virtual drop of quark
matter
\cite{Madsen1994,IidaSato1998,Lugones1998,Lugones1999,Bombaci2004}.
Which one of the two kind of droplets ($Q^*_{unp}$ or
$Q^*_{\Delta}$) will nucleate depends on the value of the
corresponding Gibbs free energy per baryon ($g_{unp}$,
$g_{\Delta}$). In fact, the latter quantity enters in the expression
of the volume term of the energy barrier separating the confined and
deconfined phases (see e.g. eq. (7) in \cite{Bombaci2004}, where the
Gibbs free energy per baryon is denoted by $\mu_i, i = Q^*, H$).
Clearly, when $g_{\Delta} < g_{unp}$  the nucleation of a
$Q^*_{\Delta}$ drop will be realized.

The {direct} formation by quantum fluctuations of a drop of
$\beta$-stable quark matter (Q phase) is also possible in principle.
However, it is strongly suppressed with respect to the formation of
the non $\beta$-stable drop by a factor $\sim G_{\mathrm{Fermi}}^{2N
/ 3}$ being $N$ the number of particles in the critical size quark
drop. This is so because the formation of a $\beta$-stable drop will
imply the almost simultaneous conversion of $\sim N/3$ {up and down}
quarks into strange quarks. For a critical size $\beta$-stable
nugget at the center of a neutron star it is found $N \sim
100-1000$, and therefore the factor is actually tiny. This is the
same reason that impedes that an iron nucleus converts into a drop
of {strange} quark matter, even in the case in which strange quark
matter had a lower energy per baryon (Bodmer-Witten-Terazawa
hypothesis). Because of this reason it is assumed that a direct
transition to $\beta$-stable quark matter is not possible
\footnote{Notice that the nucleation of an initial quark droplet
might be induced in principle by external influences such as high
energy cosmic rays or neutrinos \cite{Alcock}. However, estimates of
the production rates of quark droplets by neutrino sparking
\cite{Horvath1998} show that this mechanism is not likely to drive a
neutron to quark conversion for realistic values of the minimum
center of mass energy necessary to produce a quark-gluon plasma in
heavy ion collisions.  Ultra high energy neutrinos would be also
harmless because the outer crust acts as a shield due to the huge
cross section \cite{Horvath1998}. In this paper we are assuming that
the conversion must proceed trough an intermediate ``two flavor''
phase, but other possibilities cannot be definitely excluded.}.
Therefore, the $\beta$-stable state $Q$ could be reached only after
the $\beta$-decay of the intermediate state $Q^*$. This is in
agreement with many other previous works, see e.g.
\cite{Madsen1994,IidaSato1998,Lugones1998,Lugones1999,Bombaci2004}.

In this context, one question addressed by the present work is
whether the system settles in a paired or in an unpaired state just
after the deconfinement. On the other hand, notice that although the
above mentioned non-$\beta$-stable quark phase is very short-lived,
it constitutes an unavoidable intermediate state that must be
reached before arriving to the final $\beta$-stable configuration,
e.g. CFL quark matter (c.f. \cite{Berezhiani,Drago}). The second
question we shall address is whether this intermediate phase can
eventually preclude the transition to the final $\beta$-stable state
in spite of the latter having a lower energy. This is because the
$Q$-phase can be formed only after the nucleation of a real (i.e
critical size) drop of $Q^*_{unp}$ or $Q^*_{\Delta}$ matter, and its
subsequent ``long term'' ($t \sim  \tau_{weak} \sim 10^{-8} $s) weak
decay process.

\begin{figure}
\includegraphics[angle=0,width=7cm,clip]{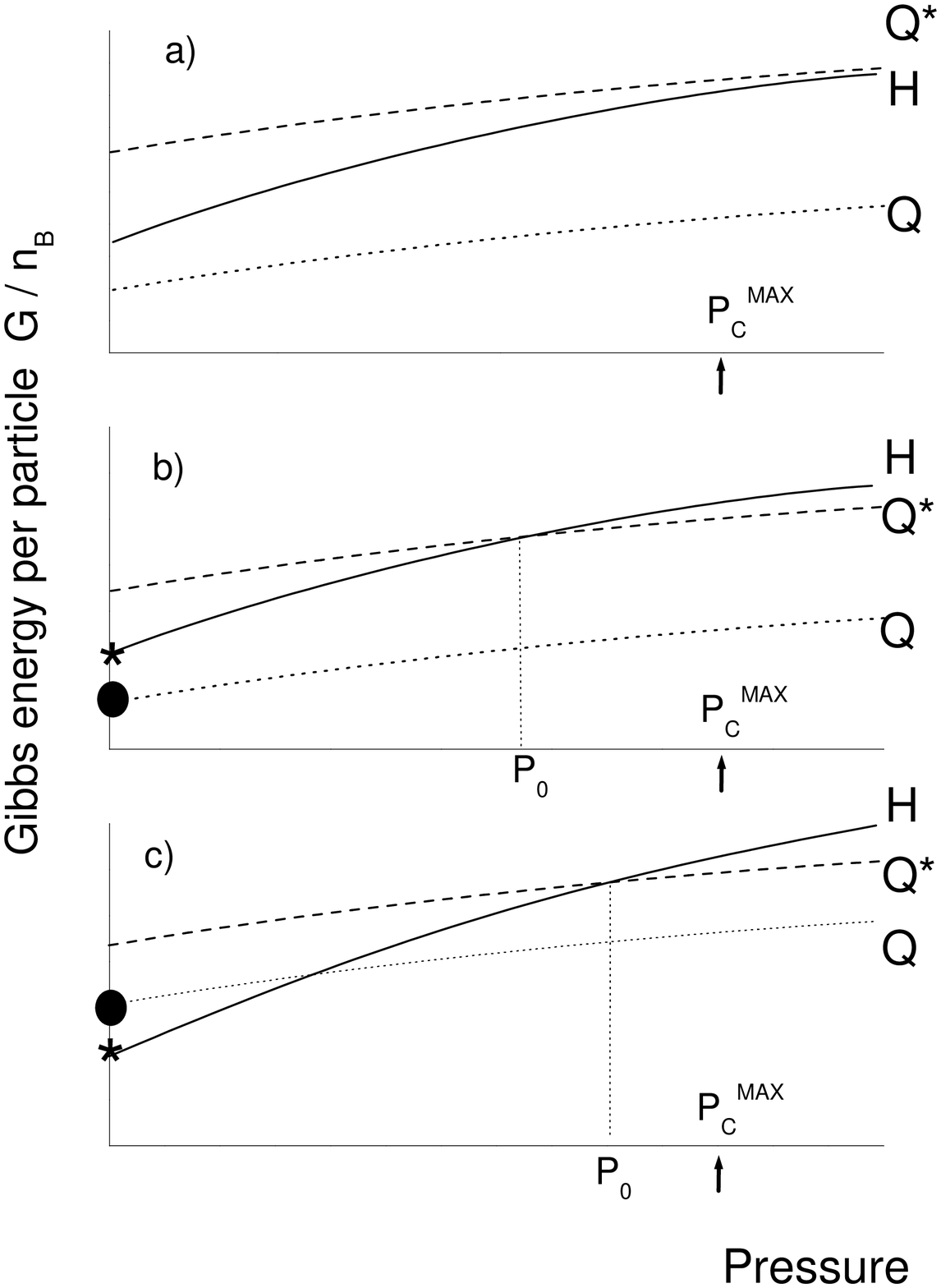}
\caption{Schematic comparison of the free energy of hadronic matter
($H$), non-$\beta$-stable "just-deconfined" matter ($Q^*$), and
$\beta$-stable quark matter ($Q$) for different cases. In the case
of panel a) the transition can never occur inside neutron stars in
spite of the final state (Q) having a lower energy per baryon. As
explained in the text, a direct transition to Q is strongly
suppressed. Since $Q^*$ has a larger energy per baryon than $H$ for
all pressures below the central pressure of the maximum mass
hadronic star ($P_c^{\mathrm{max}}$), deconfinement cannot occur
even if the $Q$ phase has a lower energy. In panels b) and c) the
phase $Q^*$ has a lower energy per baryon than $H$ for some
pressures below ($P_c^{\mathrm{max}}$). Therefore, deconfinement is
possible if pressures between $P_0$ and $P_c^{\mathrm{max}}$ are
reached inside a given neutron star. The difference between panels
b) and c) is the energy per particle at zero pressure (indicated
with a dot for $Q$ and with an asterisk for $H$).  In b) quark stars
are the so called strange stars, because they can be made up of
quark matter from the center up to the surface (P=0). In case c)
they are hybrid stars, because at zero pressure $H$ has a lower
energy than $Q$. For a fixed hadronic equation of state, these
possibilities correspond to different values of the parameters of
the quark model (the vacuum energy density $B$ and the
superconducting gap $\Delta$).} \label{fig1}
\end{figure}

\section{Deconfinement of hadronic matter into color superconducting
quark matter}

Given the uncertainties in the nature of matter at high densities,
the analysis is based on the extrapolation to higher densities of
an hadronic model valid around the nuclear saturation density
$\rho_{0}$, and the extrapolation to $\rho_{0}$ of a quark model
that is expected to be valid only for $\rho \rightarrow \infty$.
Within this kind of analysis the (in general) different functional
form of both EOSs, induces the phase transition to be first order.
Notice that from lattice QCD calculations there are indications
that the transition is actually first order in the high-density
and low-temperature regime, although this calculations involve
temperatures that are still larger than those in neutron stars,
and do not include the effect of color superconductivity
\cite{Fodor2004}.

Deconfinement is analyzed here as a first order phase transition
that conserves the flavor abundances in both phases. Therefore, it
gives a transitory colorless-quark-phase that is {\it not} in
$\beta$-equilibrium.

For describing the just deconfined quark phase we shall model it
as a free Fermi mixture of quarks and leptons and we will subtract
the pairing and the vacuum energy. The thermodynamic potential can
be written as

\begin{equation}
\Omega = \Omega_Q^{\rm free} + \Omega_Q^{\rm gap} + \Omega_L + B ,
\label{omegasc}
\end{equation}

\noindent with

\begin{equation}
\Omega_Q^{\rm free} = \sum_{f,c} \frac{1}{\pi^2} \int_{0}^{k_{fc}}
[ E_{cf} - \mu_{fc}] p^2 dp , \label{omegafree}
\end{equation}

\noindent where $E_{cf} = (p^2 + m_{fc}^2)^{1/2}$, and the sum  is
to be made over all colors and flavors of the quark mixture, being
$f=u,d,s$ the flavor index, and $c=r,g,b$ the color index.
Confinement  is introduced  in Eq. (\ref{omegasc}) by means of the
bag constant $B$, and $\Omega_L $ is the contribution of leptons.
Pairing is included through the term
\begin{equation}
\Omega_Q^{\rm gap} =  - \frac{\cal{A}}{\pi^2} \Delta^2 {\bar
\mu}^2 , \label{omegagap}
\end{equation}
which results from $E_{cf} = \sqrt{p^2 + m^2 \pm \Delta^2}$, by
expanding in Eq. (\ref{omegafree}) to
$\mathcal{O}(\Delta^2/\mu^2)$. For simplicity, all particles that
pair are assumed to have the same gap $\Delta$. The mean chemical
potential in Eq. (\ref{omegagap}) is defined through ${\bar \mu} =
N^{-1} \sum_{f,c} \mu_{fc}$ where the sum is to be made only over
particles that participate of pairing, and $N$ is the number of
different quarks that pair. The Fermi momenta are $k_{fc} = (
\mu_{fc}^2 - m_{fc}^2)^{1/2}$. The binding energy of the diquark
condensate is included by subtracting  $\Delta^2 {\bar \mu}^2 /
({4\pi^2})$ for every quasiparticle with gap $\Delta$
\cite{AfordJHEP}. As we shall see below, the relevant pairing
pattern for the deconfinement transition in neutron stars is
``2SC-like'', in which $d_r$ and $u_g$ quarks pair yielding two
quasiparticles with gap $\Delta$, and $u_r$ and $d_g$ quarks pair,
yielding two quasiparticles with gap $\Delta$ (c.f.
\cite{AfordJHEP}). Therefore, we shall set ${\cal A} = 1$ along
this work.  Note that for color-flavor-locked quark matter it is
${\cal A} = 3$.

As already emphasized by Rajagopal and Wilczeck
\cite{Rajagopal2000}, the exact nature of the interaction that
generates $\Delta$ is not relevant to the order we are working.
This means that $\Omega$ is given by the above prescription
regardless of whether the pairing is due to a point like
four-Fermi interaction, as in Nambu-Jona-Lasinio models
\cite{Buballa}, or due to the exchange of a gluon, as in QCD at
asymptotically high energies. Of course, the strength and form of
the interaction determine the value of $\Delta$, and also its
dependence with the density. Lacking of an accurate calculation
for $\Delta$, which may as large as $\sim$ 100 MeV (and even
larger in the presence of an external magnetic field
\cite{Manuel2005}), we shall keep it as a free constant parameter.

The thermodynamic quantities are straightforwardly derived from the
standard expressions: the pressure is $P = - \Omega$ and the energy
density at zero temperature is given by $\varepsilon = \sum_{Q,L}
\mu_{i}  n_{i}   + \Omega$, where the sum is to be carried over all
quarks and leptons. In particular, the particle number densities
$n_{fc} = -{\partial \Omega}/{\partial \mu_{fc}}$ are given by

\begin{equation}
n_{fc}  = \frac{k_{fc}^3}{3 \pi^2} + \frac{2 \cal{ A}}{N \pi^2}
\Delta^2 {\bar \mu}
\end{equation}
\noindent for the quarks that participate in pairing, and by

\begin{equation}
n_{fc}  = \frac{k_{fc}^3}{3 \pi^2}
\end{equation}
\noindent for quarks that do not pair. The number density of each
flavor in the quark phase is given by
\begin{eqnarray}
n_f &=& \sum_{c} n_{fc},
\end{eqnarray}
\noindent and the baryon number density $n_B$ by
\begin{equation}
n_B = \frac{1}{3} \sum_{fc} n_{fc} = \frac{1}{3} \sum_{c} n_{c} =
\frac{1}{3} \sum_{f} n_{f}.
\end{equation}

In order to close the above equations we need to impose additional
physical conditions describing the composition of the mixture
(e.g. a set of conditions on the chemical potentials). One
possibility, extensively employed in the literature, is $\beta$-equilibrium
of the quark phase. This condition describes
matter a sufficiently large time  after deconfinement ($ \tau \gg
\tau_{weak} \sim 10^{-8} s$). However, for studying deconfinement,
the relevant timescale is $\tau \ll \tau_{weak}$. As already
emphasized in
\cite{Madsen1994,IidaSato1998,Lugones1998,Lugones1999,Bombaci2004}
the appropriate condition for $\tau \ll \tau_{weak}$ is flavor
conservation between hadronic and deconfined quark matter. This
can be written as

\begin{equation}
Y^H_f = Y^Q_f   \;\;\;\;\;\; f=u,d,s,L \label{flavor}
\end{equation}

\noindent being $Y^H_f \equiv n^H_f / n^H_B$ and  $Y^Q_i \equiv
n^Q_f / n^Q_B$ the abundances of each particle in the hadron and
quark phase respectively (we shall omit the super-indexes $H$ and
$Q$ in the following). In other words, the just deconfined quark
phase must have the same ``flavor'' composition than the
$\beta$-stable hadronic phase from which it originated.

Additionally, the deconfined phase must be locally colorless;
therefore, it must be composed by an equal number or red, green
and blue quarks:

\begin{equation}
n_r = n_g = n_b  \label{colorless}
\end{equation}
\noindent being $n_r$, $n_g$ and $n_b$ the number densities of
red, green and blue quarks respectively, given by:
\begin{eqnarray}
n_c &=& \sum_{f} n_{fc}.
\end{eqnarray}
\noindent Color neutrality can be automatically fulfilled by
imposing that each flavor must be  colorless separately, i.e.
$n_{ur} = n_{ug} = n_{ub}$, $n_{dr} = n_{dg} = n_{db} $, and
$n_{sr} = n_{sg} = n_{sb}$. But in general this configuration will
not allow pairing with a significant gap. As already stated, for
quarks having different color and flavor the pairing gap may be as
large as $100$ MeV, while for particles having the same flavor the
gap is found to be about two orders of magnitude smaller (see
\cite{AfordJHEP} and references therein). Pairing is allowed even
in the case $\delta \mu > \Delta$ but the corresponding gaps are
small \cite{Huang0311155,Shovkovy2003}. Therefore, in order to
allow pairing between quarks with a non negligible gap, the Fermi
momenta of at least $u_r$ and $d_g$ quarks must be equal (the
choice of these two particular colors and flavors is just a
convention). This implies the equality of the corresponding number
densities

\begin{equation}
n_{ur} = n_{dg}. \label{2chem}
\end{equation}

\noindent The above condition represents a state that fulfills all
the physical requirements of the deconfined phase (e.g. is color
and electrically neutral), and should be the actual state (for
$\tau \ll \tau_{weak}$) if it has the lowest free energy per
baryon. Energy must be paid in order to equal at least two Fermi
seas, but in compensation the pairing energy is recovered. The
gained energy depends on the value of the pairing gap $\Delta$
and, at least for sufficiently large $\Delta$, is expected to be
larger than the energy invested to force a pairing pattern. Also,
notice that color conversion of quarks allows the adjustment of
the Fermi seas within a given flavor in a very short timescale
($\sim \tau_{strong}$), i.e. several orders of magnitude faster
than $\beta$-equilibration ($\tau_{strong} \ll \tau_{weak}$).

We emphasize that this phase  is {\it not} in flavor equilibrium.
After a weak interaction timescale this transitory pairing pattern
will be abandoned by the system in favor of the lowest-energy
$\beta$-stable configuration. Depending on the density, the lowest
energy state may be LOFF, gapless 2SC, gapless CFL, standard CFL,
(to name just some possibilities) as extensively discussed in the
literature.

\section{Application to Simplified Equations of state}

\subsection{Deconfinement of pure neutron matter}

\begin{figure}
\includegraphics[angle=-90,width=7cm,clip]{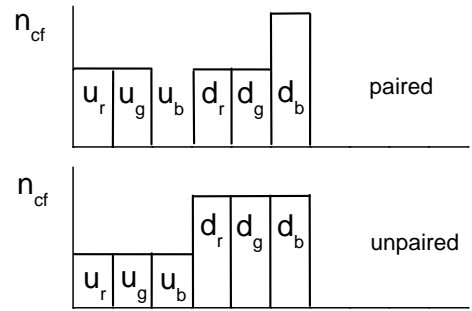}
\includegraphics[angle=-90,width=7cm]{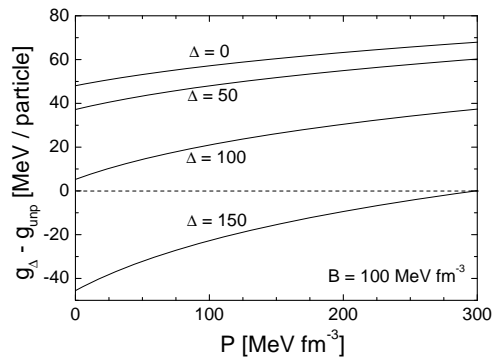}
\caption{Deconfinement of pure neutron matter. \textit{Upper panel}:
Sketch of the lowest energy configuration of the paired and unpaired
phases just after the deconfinement. \textit{Lower panel}: The
difference in the Gibbs energy per particle between paired and
unpaired quark matter. For positive values of $g_{\Delta} - g_{unp}$
the preferred phase just after the deconfinement is the unpaired one
while for negative values it is the paired one.} \label{sketch1}
\label{gibbs1}
\end{figure}

A simple solution can be found in the case of the deconfinement of
pure neutron matter, since strange quarks and electrons are not
present in the hadronic gas. First, we apply the colorless
conditions and flavor conservation introduced in the previous
section, in order determine the abundances of each quark species. In
order to allow pairing of at least two different quarks species in
the just deconfined phase,  we impose the condition of Eq.
(\ref{2chem}), i.e. $n_{ur} = n_{dg}$. Using one of the colorless
conditions ($n_r = n_g$) it is found that $n_{dr} = n_{ug}$,
implying that these quarks can also pair with a significant gap.
From the remaining colorless condition ($n_r = n_b$) it is found
$n_{ur} + n_{dr} = n_{ub} + n_{db}$. The condition of flavor
conservation states $n_d =  2 n_u$; therefore, $n_{dr} + n_{dg} +
n_{db} = 2 (n_{ur} + n_{ug} + n_{ub})$. Introducing the ratio $x =
n_{ug}/ n_{ur}$ we find from the above equations:
\begin{equation}
n_{ub}=  0  \label{}
\end{equation}
\begin{equation}
\frac{n_{db}}{n_{ur}}=  1 + x  \label{}
\end{equation}
Therefore, for massless particles at $T=0$ the chemical potentials
are related by:
\begin{eqnarray}
\mu_{ug} &=&  \mu_{dr}  = x^{1/3} \mu_{dg} = x^{1/3} \mu_{ur}  \\
\mu_{ub} &=& 0 \\
\mu_{db} &=& (1 + x)^{1/3} \mu_{ur} .
\end{eqnarray}
\noindent With this configuration, the pressure $P_{\Delta}$ and
the Gibbs free energy per baryon $g_{\Delta} = (\sum_{fc} n_{fc}
\mu_{fc} ) / n_{B}$ take the simple form
\begin{eqnarray}
P_{\Delta} &=& \frac{[2(1+ x^{4/3}) +(1+x)^{4/3}] \mu_{ur}^4}{12
\pi^2}
\nonumber \\
           & &  +\frac{(1+ x^{2/3}) \Delta^2  \mu_{ur}^2}{2 \pi^2}  - B,
\end{eqnarray}
\begin{eqnarray}
g_{\Delta}  =  \frac{[2(1+ x^{4/3}) +(1+x)^{4/3}] \mu_{ur}}{1+x}.
\end{eqnarray}
\noindent  Minimizing $g_{\Delta} = g_{\Delta}(P_{\Delta},x)$ with
respect to $x$ it is found that the minimum correspond to $x=1$.
Therefore, quarks $u_r-d_g$ and $u_g-d_r$ pair in a ``2SC-like''
pattern like the one shown in Fig. \ref{gibbs1}.

In order to determine whether the system settles in a paired or in
an unpaired configuration we compare the Gibbs free energy per
baryon of the above configuration with the Gibbs free energy per
baryon of an unpaired quark gas (both evaluated at the same
pressure, and with the same flavor composition).  For unpaired quark
matter the ground state of the colorless mixture (compatible with
flavor conservation) is shown in Fig. \ref{gibbs1}, and is described
by $n_{dr} = n_{dg} = n_{db} = 2 n_{ur} = 2 n_{ug} = 2 n_{ub} =
\frac{2}{3} n_B$. The Gibbs free energy per baryon reads
\begin{equation}
g_{unp}  = [4 \pi^2 (1+2^{4/3})^3 (P + B)]^{1/4}.
\end{equation}
The results are shown in Fig. \ref{gibbs1}  where the difference
$g_{\Delta} - g_{unp}$ is plotted as a function of pressure for
different values of the parameter $\Delta$.  For large enough values
of $\Delta$  the energy of the pairing gap is able to compensate the
increase of the free energy resulting from equating the Fermi seas
of the particles that pair. It is also worth noting that since we
are comparing the Gibbs energy per baryon of different phases at
equal pressures (alternatively, we could compare the pressure at
fixed Gibbs energy per baryon), the results depend on the bag
constant $B$.  This dependence is not found in the comparisons for
$\beta$-stable quark matter made in \cite{AfordJHEP} because the bag
constant $B$ cancels out when comparing $\Delta F$ at fixed $\mu$.
Note that $\Delta F$ is not the relevant thermodynamic function for
the analysis of the deconfinement transition in neutron stars
performed here.

\subsection{Deconfinement of $n$-$p$-$e^-$ gas}

\begin{figure}
\includegraphics[angle=-90,width=7cm,clip]{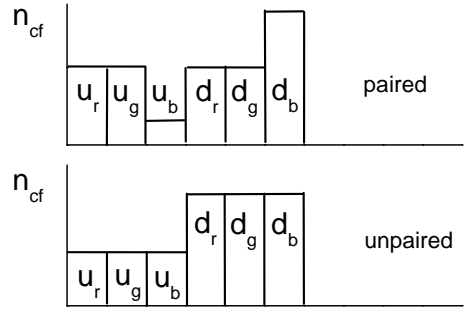}
\includegraphics[angle=-90,width=7cm,clip]{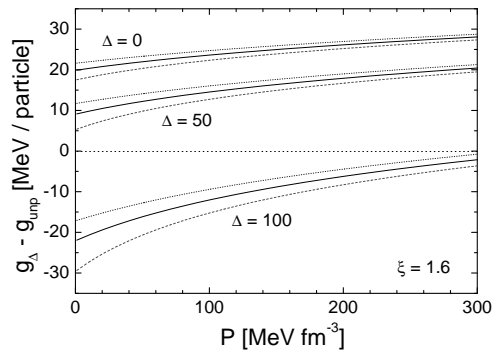}
\caption{Deconfinement of $n$-$p$-$e^-$ gas. \textit{Upper panel}:
Sketch of the lowest energy configuration of the \textit{just
deconfined} paired and unpaired phases . \textit{Lower panel}: The
difference in the Gibbs energy per baryon between both phases at
$\xi = 1.6$. For positive values of $g_{\Delta} - g_{unp}$ the
preferred phase just after the deconfinement is the unpaired one
while for negative values it is the paired one. The results are
shown for $B=60 \mathrm{~MeV~fm^{-3}}$ (dashed line), $B=100
\mathrm{~MeV~fm^{-3}}$ (solid line), and $B=140
\mathrm{~MeV~fm^{-3}}$ (doted line).} \label{gibbs2}
\end{figure}

Next we consider the deconfinement of a charge neutral uniform
system of neutrons, protons and electrons ($n_n$, $n_p$ and $n_e$
are the associated number densities). As discussed above, we impose
$n_{ur} = n_{dg}$ in order to allow pairing. Then, color charge
neutrality imposes $n_{dr} = n_{ug}$ and $n_{ur} + n_{dr} = n_{ub} +
n_{db}$. The resulting configuration allows pairing of quarks $u_r$
and $d_g$ with a gap $\Delta$, and of $u_g$ and $d_r$ with a gap
that is assumed to have the same value.

On the other hand, we can put the condition of \textit{flavor
conservation} in the following simple form

\begin{equation}
n_d = \xi n_u, \label{npe1}
\end{equation}

\noindent being $\xi \equiv Y^H_d / Y^H_u$ (c.f Eq (8)). In the case
of the $n$-$p$-$e^-$ gas, the parameter $\xi$ can be expressed in
terms of the proton fraction $Y_p = n_p / n_B$ of nuclear matter as
$\xi = (2 - Y_p)/(1 + Y_p) $.  It is easy to check that $\xi=2$
corresponds to the deconfinement of pure neutron matter, $\xi=1$ to
symmetric nuclear matter and $\xi=0.5$ to the unrealistic case of
pure proton matter. We emphasize that in the case of $\beta$-stable
$n$-$p$-$e^-$ system, $\xi$ is a function of density (or pressure)
that depends only on the state of the \textit{hadronic} matter that
deconfines.

Therefore, flavor conservation states that $n_{dr} + n_{dg} + n_{db}
= \xi (n_{ur} + n_{ug} + n_{ub})$. In addition, the condition of Eq.
(\ref{flavor}) applied to electrons yields:
\begin{equation}
3 n_{e} = 2 n_{u} - n_{d} ,\label{npe2}
\end{equation}
\noindent which confirms that flavor conservation automatically
guarantees electric charge conservation. Finally, we impose that
1) $n_{dr}= n_{ug}$ in order to allow for paring between quarks
$d_r$ with $u_g$, and 2) $n_{ur}= n_{dg}$ in order to allow for
paring between quarks $u_r$ with $d_g$.

Introducing the ratio $x = n_{ug}/ n_{ur}$, and using Eqs.
(\ref{npe1}) and (\ref{npe2}) we find the particle number
densities of each flavor and color in the paired phase as a
function of one of the particle densities (e.g. $n_{ur}$), the
parameter $\xi$ that depends only of the state of the hadronic
phase, and the free parameter $x$:
\begin{eqnarray}
n_{ug} &= & x ~n_{ur}  \\
n_{ub} &= &  (1+x) ~\frac{2-\xi}{1+\xi} ~n_{ur}  \\
n_{dr} &= & x ~n_{ur} \\
n_{dg} &= & n_{ur} \\
n_{db} &= &(1 + x) ~\frac{2 \xi - 1}{1+\xi} ~n_{ur}
\end{eqnarray}
\noindent Note that the free parameter $x$ that can be eliminated by
minimizing the Gibbs energy per baryon $g_{\Delta} = ( \sum_{fc}
n_{fc} \mu_{fc} + \mu_e n_e) /n_{B}$ with respect to $x$ at constant
pressure $P_{\Delta}$. The minimization gives $x = 1$, which means
that the configuration of the paired phase that is energetically
preferred is the one having $n_{dr}= n_{ug} = n_{ur} = n_{dg}$, as
sketched in the upper panel of Fig. \ref{gibbs1}. In Fig.
\ref{gibbs2} we compare the Gibbs energy per baryon of the
``2SC-like'' paired phase and unpaired matter for different values
of the bag constant $B$ and the pairing gap $\Delta$. Comparing with
the pure neutron matter case of Fig. \ref{gibbs1} it can be noticed
that an increase in the proton fraction of the hadronic phase favors
the formation of a paired quark phase after deconfinement.

\section{Deconfinement of cold hadronic matter}

In the following we analyze the deconfinement of a general hadronic
system including strange hadrons and then we apply the results to a
realistic EOS in order to study deconfinement inside cold neutron
stars.

\subsection{Deconfinement of a general hadronic equation of state}

\begin{figure}
\includegraphics[angle=-90,width=7cm,clip]{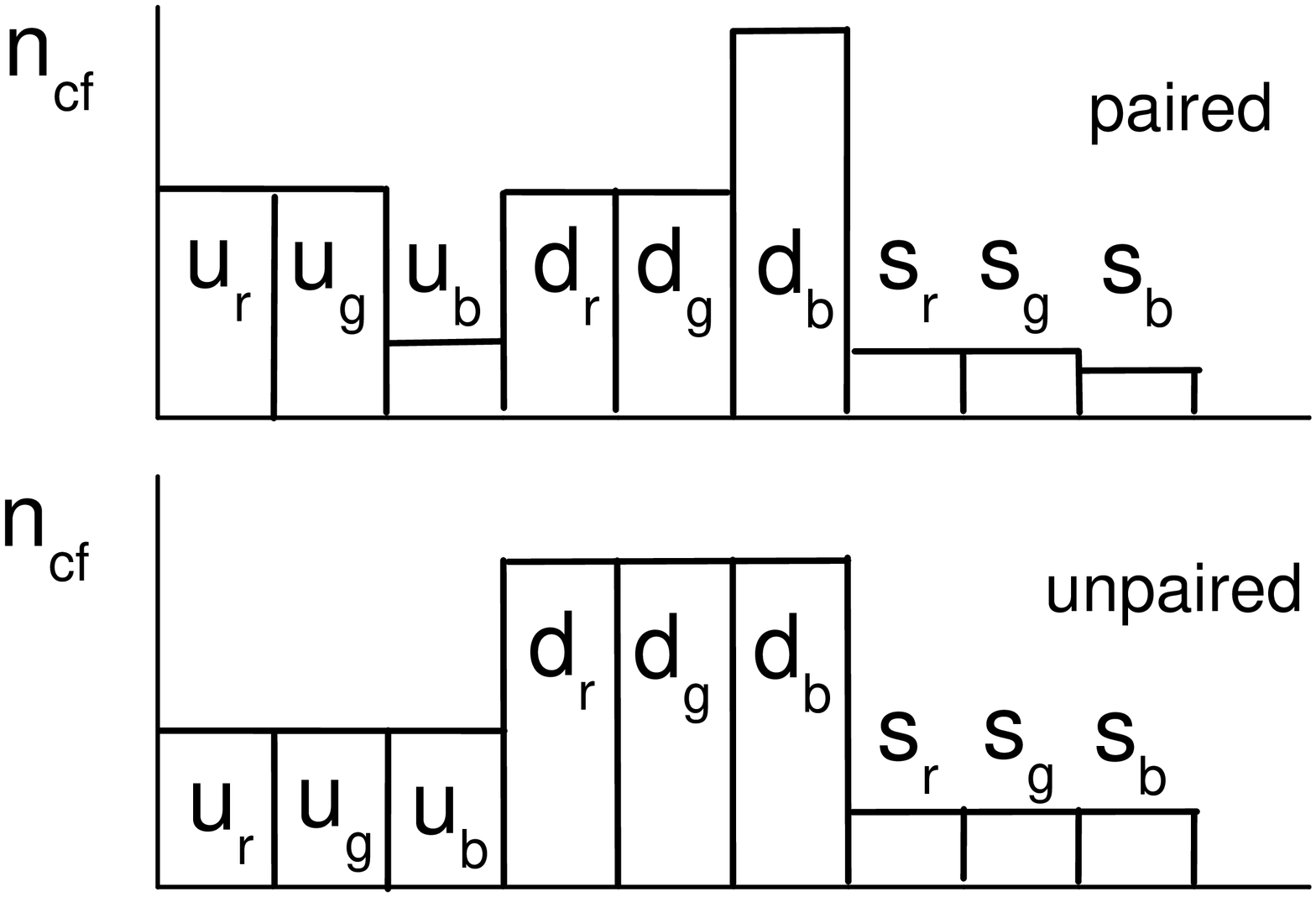}
\includegraphics[angle=-90,width=7cm,clip]{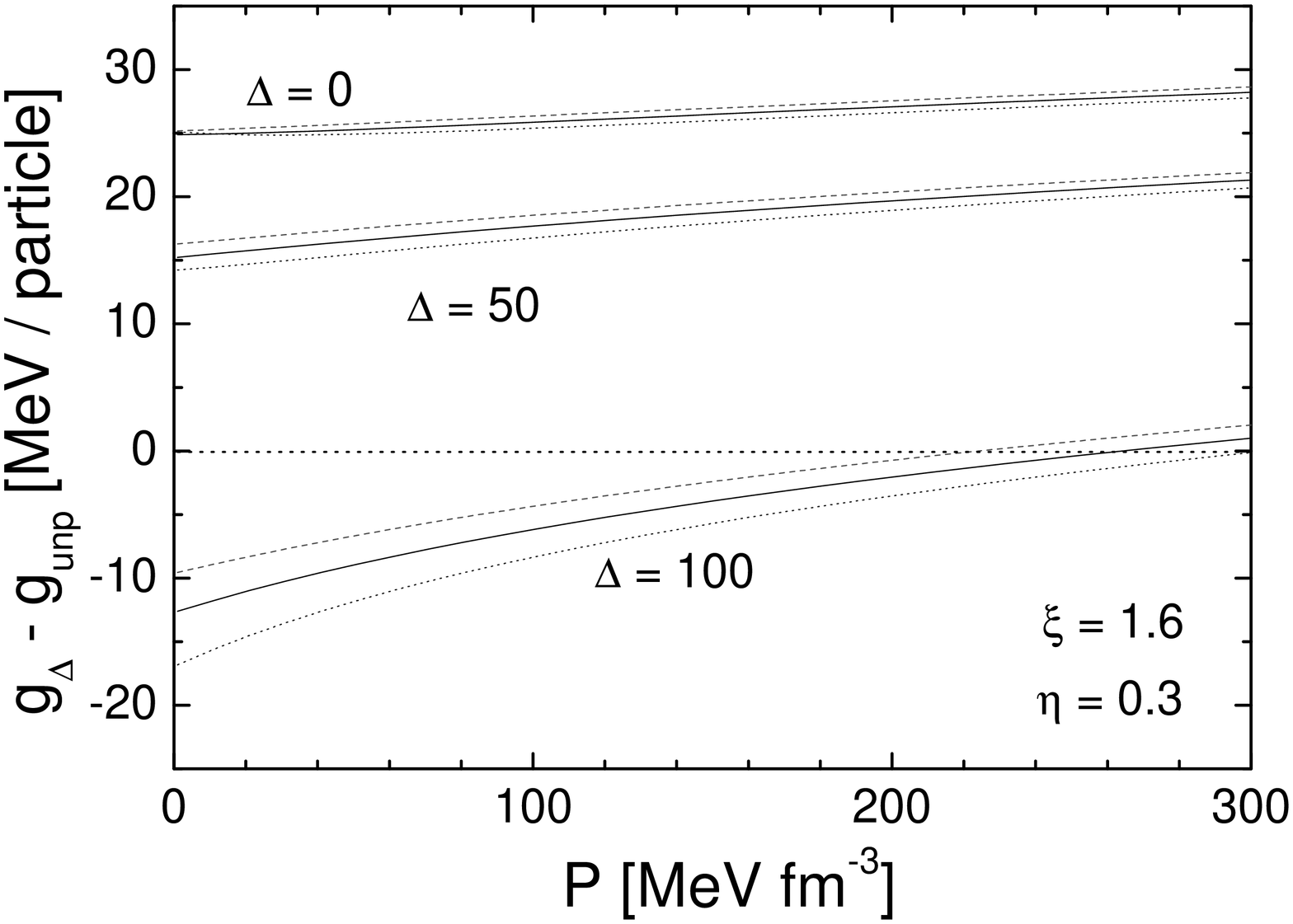}
\caption{Deconfinement of $\beta$-stable hadronic matter including
strange hadrons. \textit{Upper panel:} Sketch of the particle number
configuration just after deconfinement.  The most general paired
configuration compatible with the condition $n_{ur} = n_{dg}$,
$n_{ug} = n_{dr}$, flavor conservation, and charge neutrality is the
one sketched here. \textit{Lower panel:} The difference in the Gibbs
energy per baryon between both phases. For positive values of
$g_{\Delta} - g_{unp}$ the preferred phase just after the
deconfinement is the unpaired one while for negative values it is
the paired one.  The curves correspond to $B=60
\mathrm{~MeV~fm^{-3}}$ (dashed line), $B=100 \mathrm{~MeV~fm^{-3}}$
(solid line), and $B=140 \mathrm{~MeV~fm^{-3}}$ (doted line). We
employed $m_s = 150$ MeV.} \label{gibbs3}
\end{figure}

\textit{$\bullet$ Flavor conservation:} After deconfinement the
particle densities of quarks $u$, $d$ and $s$ are the same as in
the hadronic phase and can be determined by Eqs. (\ref{flavor}).
Another equivalent way of expressing the flavor conservation
condition is in terms of two parameters $\xi$ and $\eta$:
\begin{equation}
n_d = \xi ~ n_u. \label{h1}
\end{equation}
\begin{equation}
n_s = \eta ~ n_u. \label{h2}
\end{equation}
\noindent where $\xi \equiv Y^H_d / Y^H_u$ and $\eta \equiv Y^H_s /
Y^H_u$ depend only on the composition of the hadronic phase. These
expressions are valid for \textit{any} hadronic EOS. For hadronic
matter containing $n$, $p$, $\Lambda$, $\Sigma^{+}$, $\Sigma^{0}$,
$\Sigma^{-}$, $\Xi^{-}$, and  $\Xi^{0}$, we have
\begin{eqnarray}
\xi &=& \frac{n_p  +  2  n_n  + n_{\Lambda} + n_{\Sigma^{0}} +  2
n_{\Sigma^{-}}  + n_{\Xi^{-}}}{2  n_p  +  n_n  +  n_{\Lambda} + 2
n_{\Sigma^{+}} + n_{\Sigma^{0}}  +  n_{\Xi^{0}}}, \\
\eta &=& \frac{n_{\Lambda}  + n_{\Sigma^{+}} + n_{\Sigma^{0}}  +
n_{\Sigma^{-}} + 2 n_{\Xi^{0}} + 2 n_{\Xi^{-}}}{2  n_p  +  n_n  +
n_{\Lambda} + 2 n_{\Sigma^{+}} + n_{\Sigma^{0}}  +  n_{\Xi^{0}}}.
\end{eqnarray}
As typical values, we notice that $\eta = 0$ corresponds to zero
strangeness, and that at the center of the maximum mass star
(calculated with the hadronic equation of state of Glendenning and
Moszkowski GM1 \cite{GM1}) we have $\xi = 1.15$ and $\eta = 0.85$.
Notice that $\xi$ and $\eta$ determine univocally the number of
electrons present in the system through electric charge neutrality
of the deconfined phase:
\begin{equation}
3 n_{e} = 2 n_{u} - n_{d} - n_{s} .\label{h3}
\end{equation}

\textit{$\bullet$ Pairing condition:} As made in the previous
section for the $n$-$p$-$e^-$ gas, we impose that 1) $n_{dr}=
n_{ug}$ in order to allow for paring between quarks $d_r$ with
$u_g$, and 2) $n_{ur}= n_{dg}$ in order to allow for paring
between quarks $u_r$ with $d_g$.

\textit{$\bullet$ Color neutrality:} The condition $n_r = n_g$
leads immediately to $n_{sr} = n_{sg}$. Also, $n_r = n_b$ leads to
$ 2 n_{ur} + n_{sr} =  n_{ub} + n_{db} + n_{sb}$.

The above conditions lead to the pairing pattern schematically shown
in the upper panel of Fig. \ref{gibbs3}. Note that these conditions
still leave a degree of freedom that can be fixed by introducing an
additional parameter $h$ relating the particle number densities of
two arbitrary quark species. Therefore, it is possible to impose the
equality of two arbitrary Fermi seas in order to allow pairing
between them. We have analyzed the 10 possible combinations and
verified that 6 of them lead to a negative value of the particle
number density of at least one quark species. The other 4
possibilities allow pairing of particles that don't have different
color and flavor, allowing pairing with a negligible gap. For this
reason, it is more convenient to introduce  $h \equiv
n_{sb}/n_{sr}$, and minimize the free energy with respect to $h$.

Using the above Eqs. we find the following linear set of
equations:
\begin{eqnarray}
2 n_{ur} + n_{sr} & = &  n_{ub} + n_{db} + n_{sb} \\
2 n_{ur} + n_{db} & = & \xi (2 n_{ur} + n_{ub}) \\
2 n_{sr} + h ~ n_{sr} & = & \eta (2 n_{ur} + n_{ub}),
\end{eqnarray}
\noindent from which we obtain the number densities of each quark
species in the paired phase as functions of only four quantities:
\begin{eqnarray}
{n_{ub}} & =& 2\frac{4 + \eta + 2h - \eta h - 2 \xi - h \xi}{2 -
\eta + h + \eta h + 2 \xi + h \xi} ~{n_{ur}} \label{n1}\\
{n_{db}} & = & 2\frac{-2 + \eta - h - \eta h + 4 \xi + 2 h \xi}{2
- \eta + h + \eta h + 2 \xi + h \xi} ~{n_{ur}} \label{n2}\\
{n_{sb}} & = & \frac{6 \eta h}{2 - \eta + h + \eta h + 2 \xi + h
\xi} ~{n_{ur}} \label{n3}\\
{n_{sr}} &=& \frac{6 \eta}{2 - \eta + h + \eta h + 2 \xi + h \xi}
~{n_{ur}} \label{n4}\\
{n_{e}} & = & \frac{2 (2 + h) (2 - \eta - \xi )}{2 - \eta + h +
\eta h + 2 \xi + h \xi} ~{n_{ur}}.   \label{n5}
\end{eqnarray}

\noindent Remind that the other particle densities are given by
$n_{ug} = n_{dr} = n_{dg} = n_{ur}$, $n_{sg} = n_{sr}$, and
$n_{sb} = h n_{sr}$.

The pressure and Gibbs energy per baryon of the paired deconfined
phase can also be written in terms of the same parameters:
\begin{eqnarray}
P_{\Delta} &=& \sum_{fc} \frac{k^4_{fc}}{12 \pi^2}  +
\frac{\mu^4_{e}}{12 \pi^2}  + \frac{1}{\pi^2} \bar{\mu}^2
\Delta^2 - B,  \label{eosq1}\\
g_{\Delta}  &=& \sum_{fc} \frac{n_{fc} \mu_{fc}}{n_B}  + \frac{\mu_e
n_e}{n_{B}},
\end{eqnarray}
\noindent where $k_{fc} = (\mu^2_{fc} - m^2_{fc})^{1/2}$,
$\bar{\mu} = \mu_{ur}$,  the chemical potentials $\mu_{fc}$ are
obtained from
\begin{eqnarray}
n_{fc} = \frac{\mu_{fc}^3}{3 \pi^2} + \frac{2 \Delta^2
\bar{\mu}}{\pi^2} & & f_c = u_r, u_g, d_r, d_g  \\
\mu_{fc} = (3 \pi^2 n_{fc})^{1/3} & & f_c = u_b, d_b \\
\mu_{fc} = [(3 \pi^2 n_{fc})^{2/3} + m_s^2]^{1/2} & & f_c = s_r, s_g, s_b
\end{eqnarray}

The minimization of $g_{\Delta}$ with respect to $h$ gives $h=1$ and
therefore the number densities are given by the following equations:

\begin{eqnarray}
{n_{ub}} & =& \frac{ 4 - 2\xi}{1  +  \xi} ~ {n_{ur}} \label{1}\\
{n_{db}} & = & \frac{ 4 \xi - 2}{1 + \xi } ~{n_{ur}} \label{2}\\
{n_{sb}} & = & \frac{2 \eta }{1  + \xi} ~{n_{ur}} \label{3}\\
{n_{e}} & = & \frac{2 (2 - \eta - \xi )}{1 + \xi} ~{n_{ur}}.
\label{eosq10}
\end{eqnarray}

\noindent with $n_{ug} = n_{dr} = n_{dg} = n_{ur}$ and  $n_{sg} =
n_{sr}= n_{sb}$. Equations (\ref{eosq1})-(\ref{eosq10}) constitute
the equations of state for just deconfined quark matter. In the
lower panel of Fig. \ref{gibbs3} we show $\Delta g$ for particular
values of the parameters ($\xi = 1.6$, $\eta = 0.3$).

\subsection{Deconfinement transition in neutron stars}

\begin{figure}
\includegraphics[angle=-90,width=9cm,clip]{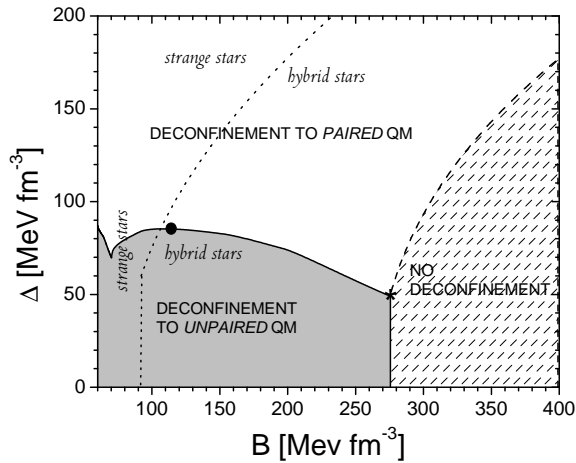}
\caption{The cparameter space $\Delta$ vs. $B$ indicating the
regions for which deconfinement is possible inside the maximum mass
neutron star with the GM1 EOS \cite{GM1} ($M_{max} = 1.8
M_{\odot}$). We also indicate whether the final state reached after
$\beta$-equilibration of the just-deconfined phase has energy per
baryon less or greater than the neutron mass (i.e. leads to the
formation of  strange stars or hybrid stars respectively). We
adopted $m_s = 150$ MeV for the strange quark mass. If $(B,\Delta)$
fall inside the dashed region, deconfinement is not possible even at
the center of the maximum-mass star with this EOS. For $(B,\Delta)$
inside the grey region the just-deconfined unpaired phase has always
less energy per baryon than the just-deconfined paired phase. For
$(B,\Delta)$ in the white region the just-deconfined phase is always
paired quark matter. The regions met at a point of coordinates
($B_*,\Delta_*$) indicated with an asterisk and shown in Table I for
different values of the strange quark mass $m_s$. The maximum of the
grey region is indicated with a dot, and the corresponding value
$\Delta_{max}$ is shown in Table I for different $m_s$.}
\label{Pstar1}
\end{figure}

\begin{table}
\begin{tabular}{cccc}
\hline
$m_s$ (MeV)~~& $B_*$ (MeV fm$^{-3}$)~~ & $\Delta_*$ (MeV) & $\Delta_{max}$ (MeV)\\
\hline
0 &  323 & 34 & 78\\
100 & 300 & 37 & 80 \\
150 &  275 &  49 & 85 \\
200 &  241  &  68 &  97 \\
\hline
\end{tabular}
\caption{We show the coordinates of the point ($B_*,\Delta_*$)
(indicated with an asterisk in Fig. \ref{Pstar1}) for different
values of the strange quark mass $m_s$. We also list the coordinate
$\Delta_{max}$ of the point indicated with a dot in Fig.
\ref{Pstar1}. The qualitative shape of Fig. \ref{Pstar1} remains the
same in all cases, but the size of each region changes according
with the representative points given here.} \label{table}
\end{table}

\begin{figure}
\includegraphics[angle=-90,width=9cm,clip]{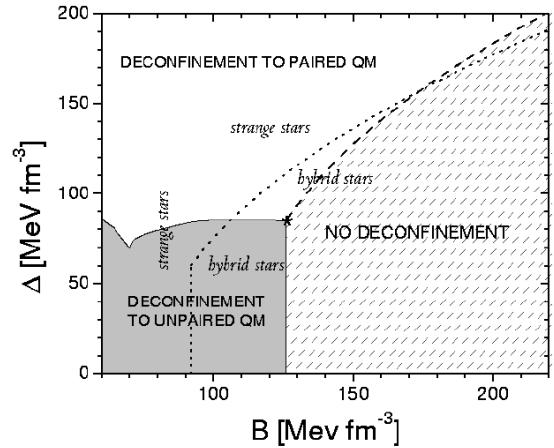}
\caption{The same as Fig. \ref{Pstar1} but for a 1.6 $M_{\odot}$
neutron star. The solid curve is the same as in Fig. \ref{Pstar1}
but the dashed curve is strongly shifted to the left, i.e., the
region of the parameter space that allows deconfinement for a 1.6
$M_{\odot}$ neutron star is much smaller than for a 1.8 $M_{\odot}$
neutron star. Notice that the larger part of the available parameter
space would lead to the formation of strange stars rather than
hybrid stars. } \label{Pstar2}
\end{figure}

The above conditions allow us to study the regions of the
parameter space $\Delta$  versus $B$ where the deconfinement
transition is possible inside neutron stars (see Figs.
\ref{Pstar1} and \ref{Pstar2}). In this analysis we shall employ
the equations of state of Glendenning and Moszkowski GM1 for the
hadronic phase \cite{GM1}. Depending on the value of $B$ and
$\Delta$ there are three possibilities: 1) deconfinement is not
possible at the center of the neutron star, 2) deconfinement to an
unpaired phase is preferred, and 3) deconfinement to a paired
phase is preferred. These regions are limited by the following
curves:

\begin{itemize}
\item A curve along which the three phases have the same state,
that is, $g_H(P,\eta,\xi) = g_{unp}(P,\eta,\xi) =
g_{\Delta}(P,\eta,\xi)$ evaluated at the same pressure $P_H =
P_{unp} = P_{\Delta}$ and with the same ``quark composition'', i.e.
$\eta_H = \eta_{unp} = \eta_{\Delta}$ and $\xi_H = \xi_{unp} =
\xi_{\Delta}$. This gives the solid line in Figs. \ref{Pstar1} and
\ref{Pstar2} separating the grey and white regions.

\item A curve along which $g$, $P$, $\xi$ and $\eta$ of
quark matter has the same value than at the center of the hadronic
neutron star (in Figs. \ref{Pstar1} and \ref{Pstar2}: dashed curve
for paired quark matter and solid vertical line for unpaired quark
matter). The position of this curve depends on the mass of the
neutron star. In Fig. \ref{Pstar1} we show the results for the
maximum-mass hadronic star within the GM1 EOS (1.8 $M_{\odot}$), and
in Fig. \ref{Pstar2} for a neutron star with 1.6 $M_{\odot}$.
\end{itemize}

The regions meet at a point of coordinates ($B_*,\Delta_*$)
indicated with an asterisk in Figs. \ref{Pstar1} and \ref{Pstar2}.
The position of this point (that characterizes rather well the size
of the regions) depends on the assumed value of the strange quark
mass $m_s$, and on the mass of the neutron star. In Table I we show
the dependence of ($B_*,\Delta_*$) on the assumed value of the
strange quark mass $m_s$ for the 1.8 $M_{\odot}$ neutron star of Fig
\ref{Pstar1}. We also indicate with a dot the maximum of the grey
region (which is the same in Figs. \ref{Pstar1} and \ref{Pstar2})
and give the corresponding value $\Delta_{max}$ in Table I.

We have also included in the parameter space the curve separating
the regions in which $\beta$-stable quark matter has an energy per
baryon smaller than the neutron mass from the region in which
$\epsilon/n_B(P=0) > m_n $ (for simplicity, paired $\beta$-stable
quark matter is assumed in all cases to be CFL). To the left of this
curve the final state after $\beta$-equilibration is absolutely
stable quark matter leading to the formation of strange stars. To
the right, $\beta$-stable quark matter is restricted to the core of
neutron stars (hybrid stars). The position of this curve also
depends on the value of $m_s$. In Figs. \ref{Pstar1} and
\ref{Pstar2} it is shown for $m_s = 150$ MeV (for more details the
reader is refereed to \cite{Lugones2002}).

\section{Discussion}

In this paper we have analyzed the deconfinement transition from
hadronic matter to quark matter, and investigated the role of color
superconductivity in this process. We have deduced the equations
governing deconfinement when quark pairing is allowed and, employing
a realistic equation of state for hadronic matter, we have found the
regions of the parameter space $B$ versus $\Delta$ where the
deconfinement transition is possible inside neutron stars. The main
results are shown in Figs. \ref{Pstar1} and \ref{Pstar2} and were
explained in the last section. In the following we discuss some
implications for neutron star structure.

Stars containing quark phases fall into two main classes: hybrid
stars (where quark matter is restricted to the core) and strange
stars (made up completely by quark matter). This structural
characteristic depends on whether the energy per baryon of
$\beta$-equilibrated quark matter at zero pressure and zero
temperature is less than the neutron mass (the so called ``absolute
stability'' condition). In the absence of pairing, quark matter in
$\beta$-equilibrium has an energy per baryon (at $P=0$) smaller than
the neutron mass only if $B$ is in the range $57$ MeV fm$^{-3}$ $
\lesssim B \lesssim$  90 MeV fm$^{-3}$. Within this range of $B$,
unpaired $\beta$-stable quark matter is the so called strange quark
matter, and it is possible the existence of stars made up entirely
by the quark phase. For $B \gtrsim 90$ MeV fm$^{-3}$ unpaired
$\beta$-stable quark matter at $P=0$ and $T=0$ decays into hadrons,
and therefore it can be present only in the core of neutron stars.
The size of the core (if any) depends on the value of $B$: the
larger the value of $B$, the smaller the size of the quark matter
core (for a given neutron star mass).

Pairing enlarges substantially the region of the parameter space
where $\beta$-stable quark matter has an energy per baryon smaller
than the neutron mass \cite{Lugones2002,Lugones2003}. Although the
gap effect does not dominate the energetics, being of the order
($\Delta / \mu)^2 \sim$ a few percent, the effect is substantially
large near the zero-pressure point (which determines the stability
and also the properties of the outer layers and surface of the
star). As a consequence, a ``CFL strange matter'' is allowed for the
same parameters that would otherwise produce unbound strange matter
without pairing \cite{Lugones2002}. The line separating strange
matter from non-absolutely stable quark matter is shown in dotted
line in Figs. \ref{Pstar1} and \ref{Pstar2}, according to
\cite{Lugones2002}.

Concerning just deconfined quark matter (i.e. not in
$\beta$-equilibrium) it has been already shown that the transition
to \textit{unpaired} quark matter is not possible in a 1.6
$M_{\odot}$ neutron star if the Bag constant is $B \gtrsim 126$ MeV
fm$^{-3}$, because the transition pressure is never reached inside the
star, even in the proto-neutron star phase \cite{Lugones1999}. The
results when pairing is allowed have been shown in the previous
section, where we have shown the ``deconfinement'' parameter space
for the maximum mass neutron star with the GM1 EOS (1.8
$M_{\odot}$), and for a 1.6 $M_{\odot}$ neutron star. As it is
evident from Figs. \ref{Pstar1} and \ref{Pstar2}, deconfinement is
facilitated for large $\Delta$ (i.e. it is possible for a larger
range of $B$). This result can be roughly understood if we think
paired matter as unpaired matter with an effective bag constant
depending on the chemical potential (or on density):
$B_{\mathrm{eff}}(\Delta,\mu) = B - \frac{\cal{A}}{\pi^2} \Delta^2
{\bar \mu}^2$. The minus sign of the condensation term in the
previous expression allows deconfinement for larger values of $B$.
Nevertheless, notice that this simple interpretation is not strictly
correct because the chemical equilibrium is different in paired and
unpaired phases. In fact, due to the more convenient chemical
equilibrium, the unpaired phase is energetically preferred for small
$\Delta$.

For a large part of the deconfined parameter-space $\Delta$ vs. $B$,
$\beta$-stable quark matter has (at any pressure) an energy per
baryon smaller than the neutron mass. The situation corresponds to
the one sketched in Fig. 1b. Therefore, if quark stars are formed,
they would be made up of quark matter from the center up to the
surface. In Figs. \ref{Pstar1} and \ref{Pstar2} this corresponds to
the part of the grey and the white regions to the left of the dotted
line.

To the right of the dotted line of Figs. \ref{Pstar1} and
\ref{Pstar2} stars containing quark phases are hybrid. The situation
corresponds to the one sketched in Fig. 1c, i.e. the hadronic phase
is preferred at low pressures. From the point of view of the
structural properties, stable hybrid stars have been found to be
possible in a wide region of the parameter space
\cite{Alford-Reddy}. Nevertheless, notice that these configurations
could not form in practice if they fall in the region where
deconfinement is forbidden (dashed region of Figs. \ref{Pstar1} and
\ref{Pstar2}).
That is, for a given hadronic star, there exist a stable hybrid star
with the same baryonic mass that has a lower energy {(gravitational
mass)}. Nevertheless, the hadronic star cannot deconfine because the
here studied non-$\beta$-stable \emph{intermediate} state has a
larger energy per baryon, as shown schematically in Fig. 1a. This is
the case, for example, if $B$ = 153 MeV fm$^{-3}$ ($B^{1/4} = 185$
MeV) as can be seen by comparing with the results of Fig. 5 of the
paper by Alford and Reddy \cite{Alford-Reddy}: for $m_s = 150$ MeV,
$B$ = 153 MeV fm$^{-3}$ and any reasonable value of $\Delta$, stable
hybrid configurations with maximum masses up to 1.6 $M_{\odot}$ are
found in \cite{Alford-Reddy}. However, in these cases the pairs
($B,\Delta$) fall comfortably inside the dashed region of Fig.
\ref{Pstar2} where deconfinement is not allowed. For the same value
of $B$, heavier stars ($\sim$ 1.8 $M_{\odot}$) could deconfine,
since ($B,\Delta$) would be inside the grey region of Fig.
\ref{Pstar1}, but the resulting configuration would be not
structurally stable and would form a black hole (c.f.
\cite{Alford-Reddy}). Although the EOSs are different in
\cite{Alford-Reddy} and in the present work, this should not affect
this generic trend. Notice that qualitatively similar results have
been found in \cite{Lugones1999,Bombaci2004} for unpaired quark
matter.

A stated in the Introduction, the transition from nuclear matter to
quark matter proceeds by bubble nucleation. However, notice that for
large $B$ the results with typical surface tension $\sigma =
\mathrm{10- 30 ~ MeV fm^{-2}}$ do not differ much from the case in
bulk \cite{IidaSato1998,Bombaci2004}. This means that we don't
expect that the dashed region of Figs. \ref{Pstar1} and \ref{Pstar2}
will change significantly when including surface effects. Anyway,
even if the surface tension were very large, the here presented bulk
case is still relevant because it gives a lower limit for the
transition: i.e., if deconfinement is not possible in bulk, it will
be even more difficult when including surface effects. In other
words, the dashed line of Figs. \ref{Pstar1} and \ref{Pstar2} could
move to the left in a more refined study, but not to the right. A
complete study of the astrophysical implications is in progress and
will be published elsewhere.

\section{Acknowledgements}

G.L. wants to thank FAPESP for support during an early phase of this work.
We thank Jorge Horvath and Ettore Vicari for stimulating discussions.

\end{document}